\documentclass[aps,prd,preprint,showpacs,preprintnumbers]{revtex4}
\usepackage{graphics}
\def \beq{\begin{equation}}
\def \eeq{\end{equation}}
\def \beqa{\begin{eqnarray}}
\def \eeqa{\end{eqnarray}}

\def \O{{\cal O}}
\def \ie{{\sl i.e.\/}}
\def \etc{{\sl etc.\/}}
\def \tr{{\rm Tr}\,}
\def \det{{\rm Det}\,}
\begin{document}

\title{Pressure and non-linear susceptibilities in QCD at finite chemical potentials}
\author{Rajiv V.\ \surname{Gavai}}
\email{gavai@tifr.res.in}
\author{Sourendu \surname{Gupta}}
\email{sgupta@tifr.res.in}
\affiliation{Department of Theoretical Physics, Tata Institute of Fundamental
         Research,\\ Homi Bhabha Road, Mumbai 400005, India.}

\begin{abstract}
When the free energy density of QCD is expanded in a Taylor series in the
chemical potential, $\mu$, the coefficients are the non-linear
quark number susceptibilities.  We show that these depend on the
prescription for putting chemical potential on the lattice, making
all extrapolations in chemical potential prescription dependent at
finite lattice spacing.  To put bounds on the prescription dependence,
we investigate the magnitude of the non-linear susceptibilities over a
range of temperature, $T$, in QCD with two degenerate flavours of light
dynamical quarks at lattice spacing $1/4T$. The prescription dependence
is removed in quenched QCD through a continuum extrapolation, and the
dependence of the pressure, $P$, on $\mu$ is obtained.
\end{abstract}
\pacs{11.15.Ha, 12.38.Gc\hfill TIFR/TH/03-07, hep-lat/0303013}
\preprint{TIFR/TH/03-07, hep-lat/0303013}
\maketitle

One of the most important objects in the study of hot and dense
hadronic matter is the phase diagram, particularly, the location of
the critical end point, characterised by the temperature $T_E$ and the
chemical potential $\mu_E$.  Much effort has been expended recently
on estimating these quantities at finite lattice spacing, $a$, using,
implicitly \cite{fodor} or explicitly \cite{biswa,mpl,pop}, a Taylor
series expansion of the free energy density. This needs the non-linear
susceptibilities which define the response to an applied $\mu$ beyond
quadratic order. An equally important question for phenomenology arises
from the fact that present day heavy-ion collision experiments access
the part of the QCD phase diagram with $\mu\simeq$10--80 MeV, \ie,
baryon chemical potential $\mu_B\simeq$30--250 MeV \cite{cleymans},
far from $\mu_E$.  It is then pertinent to ask how relevant the $\mu=0$
lattice QCD computations of quantities such as the pressure, $P$, are
to these experiments.

In this paper we present the first investigation of these non-linear
susceptibilities. We uncover essential lattice artifacts, but manage to
quantify and remove them in the process of taking the continuum limit.
We explicitly construct a Taylor series expansion for $P$ at $\mu>0$, put
limits on the region of linear response, \ie, of reliable extrapolations,
and show that the $\mu=0$ lattice computations are clearly relevant
to experiments.  An interesting sidelight is that there is strong
evidence of short thermalisation times in the dense matter formed in
these heavy-ion collisions \cite{heinz}, which may be related to large
values of transport coefficients \cite{gupta}. Most computations of such
dynamical quantities are based on linear response theory. The success of
the linear approximation in static quantities at fairly large driving
also gives us confidence in using linear response theory for dynamics.
Another interesting point is that the radius of convergence of a Taylor
series expansion started near $T_c$ \cite{expltc} must give information
on the location of the critical end-point, $(T_E,\mu_E)$,  through the
Taylor coefficients, \ie, the non-linear susceptibilities. Since these
Taylor coefficients turn out to be prescription dependent and subject
to strong finite lattice spacing effects, it seems that present day
estimates of the end point will have to be sharpened strongly before
they can be used as a guide to phenomenology.

\begin{figure}[hbt]\begin{center}
   \scalebox{0.5}{\includegraphics{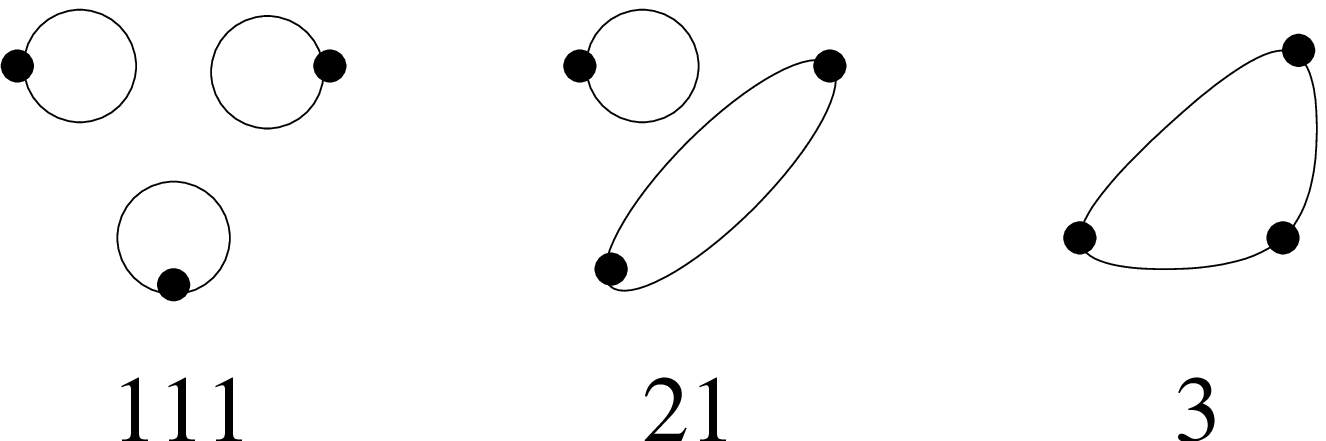}}
   \scalebox{0.5}{\includegraphics{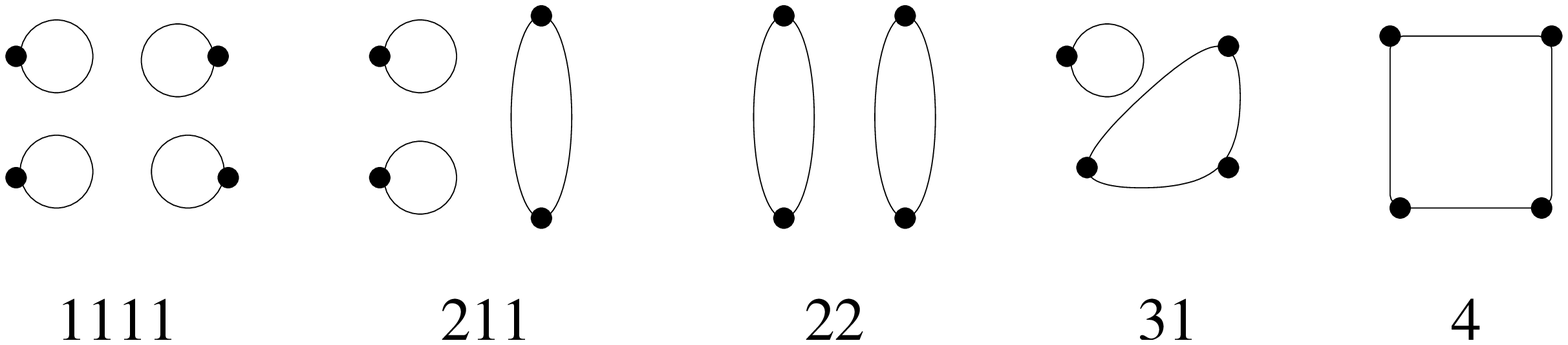}}
   \end{center}\vskip-5mm
   \caption{All topologies which contribute to derivatives up to fourth
      order, and the notation for the corresponding operators.}
\label{fg.ops}\end{figure}

The partition function of QCD at finite temperature $T$ and chemical
potentials $\mu_f$ for each flavour $f$ can be written as
\beq
   Z \equiv {\rm e}^{-F/T} = \int{\cal D}U{\rm e}^{-S(T)}
         \prod_f\det M(m_f,T,\mu_f).
\label{part}\eeq
$F$ is the free energy, $S$ is the gluon part of the action, $M$ is
the Dirac operator, each determinant is for one quark flavour and the
temperature $T$ enters through the shape of the lattice and boundary
conditions \cite{footz}.  We shall work with a lattice discretisation
and use staggered quarks \cite{footdet}.  In this work we shall only
consider two degenerate flavours of quarks--- $m_u=m_d=m$ \cite{nf11},
with chemical potentials $\mu_u$ and $\mu_d$.  The number densities,
$n_f$, and the (linear) quark number susceptibilities, $\chi_{fg}$,
are the first and second derivatives of $-F/V$ with respect to $\mu_f$
and $\mu_g$ \cite{gott}. Since $P=-F/V$ for a homogeneous system, the
non-linear susceptibilities of order $n\ge3$ are also the remaining
Taylor coefficients of an expansion of $P$---
\beq
   \chi_{fg\cdots} = 
      -\frac1V\frac{\partial^n F}{\partial\mu_f\partial\mu_g\cdots} =
       \frac TV\frac{\partial^n \log Z}{\partial\mu_f\partial\mu_g\cdots},
\label{nlins}\eeq
where we construct the expansion around $\mu_f=0$.

We now write systematic rules for the construction of the non-linear
susceptibilities.  The derivatives of $\log Z$ needed in eq.\
(\ref{nlins}) can be related to the derivatives of $Z$ with respect to
the chemical potentials $\mu_f$, $\mu_g$, \etc, (which we denote by
$Z_{fg\cdots}$) by the usual formul{\ae} for taking connected parts
\cite{momcum}. The only extra point to remember is that all the odd
derivatives vanish by CP symmetry. To write the subsequent formul{\ae}
compactly, we define operators $\O_i$ by
\beq
   Z_f=Z\langle\O_1\rangle, \qquad{\rm and}\qquad
   \O_{n+1} = \frac{\partial\O_n}{\partial\mu_f},
\label{defo}\eeq
where angular brackets denote averages over the ensemble defined by eq.\
(\ref{part}) at $\mu_f=0$.  Diagrammatic rules \cite{zak} for the $\O_i$
and the derivatives of $Z$, are--
\begin{enumerate}
\item Put down $n$ vertices (each corresponding to a derivative of $M$
   with respect to $\mu_f$) and label each with its flavour index.
\item Join the vertices by lines (each representing a quark) into sets of
   closed loops such that each loop contains only vertices of a single flavour.
   $\O_i$ is denoted by a single loop joining $i$ vertices.
\item For degenerate flavours and $\mu_f=0$, the operators are labeled only by
   the topology, which is specified completely by the number of vertices per
   loop and the number of such loops. Therefore erase the flavour index after
   step 2. We denote each resulting operator by the notation
   $\O_{ij\cdots}=\O_i\O_j\cdots$, where $i+j+\cdots=n$.
\item For each $n$-th order derivative of $Z$, add all the operator topologies
   for fixed $n$ with flavour-dependent multiplicity equal to the number of ways
   in which each topology arises given the flavour indices.
\end{enumerate}
The number densities $n_u=n_d=(T/V)\langle\O_1\rangle$ vanish
at $\mu=0$. We have considered the (linear) susceptibilities
$\chi_3=(T/V)\langle\O_2\rangle$ and $\chi_{ud}=(T/V)\langle \O_{11}
\rangle$ extensively in a recent series of papers \cite{conti}.
The new quantities that we now consider are the two third order
derivatives 
\beq
   Z_{uuu} = Z\langle\O_3 + 3\O_{12} + \O_{111}\rangle
    \qquad{\rm and}\qquad
   Z_{uud} = Z\langle\O_{12} + \O_{111}\rangle,
\label{ord3}\eeq
the three fourth order derivatives 
\beqa
\nonumber
   Z_{uuuu} &=& Z\langle\O_4 + 4\O_{13} + 3\O_{22} + 6\O_{112} + \O_{1111}\rangle,\\
\nonumber
   Z_{uuud} &=& Z\langle\O_{13} + 3\O_{112} + \O_{1111}\rangle,\\
   Z_{uudd} &=& Z\langle\O_{22} + 2\O_{112} + \O_{1111}\rangle,
\label{ord4}\eeqa
and the five corresponding susceptibilities. The third order susceptibilities
turn out to vanish.  The fourth order susceptibilities are
\beqa
\nonumber
   \chi_{uuuu} &=& \left(\frac TV\right)
       \left[\frac{Z_{uuuu}}Z-3\left(\frac{Z_{uu}}Z\right)^2\right],\\
\nonumber
   \chi_{uuud} &=& \left(\frac TV\right) \left[\frac{Z_{uuud}}Z
       -3\left(\frac{Z_{uu}}Z\right)\,\left(\frac{Z_{ud}}Z\right)\right],\\
   \chi_{uudd} &=& \left(\frac TV\right)
       \left[\frac{Z_{uudd}}Z-\left(\frac{Z_{uu}}Z\right)^2
               -2\left(\frac{Z_{ud}}Z\right)^2\right].
\label{chi4}\eeqa
The operators contributing to eqs.\ (\ref{ord3}, \ref{ord4}) are shown
in Figure \ref{fg.ops}.  Note the interesting fact that beyond the
second order, the number of distinct operator topologies is greater
than the number of susceptibilities \cite{zak}; however by making $N_f$
sufficiently large, all topologies up to any given order can be given
a physical meaning.

A perturbative expansion in the continuum proceeds through an
order-by-order enumeration of interaction terms.  In the continuum
the diagrams in Figure \ref{fg.ops} are the leading order (ideal quark
gas) part of the perturbative expansion of the susceptibilities, where
each vertex corresponds to the insertion of a $\gamma_0$ (since
the chemical potential enters the Lagrangian as $\gamma_0\mu_f$).
Higher order Feynman diagrams correspond to dressing these loops by
gluon attachments in all possible ways.

In the lattice theory the diagrams in Figure \ref{fg.ops} stand for
operator definitions which need further specification.  They are not
Feynman diagrams, but mnemonics
for the process of taking derivatives of $Z$.  Since, the coupling of
Fermions to the chemical potential is non-linear \cite{chempot}, hence
all derivatives of $M$ exist and are non-zero in general.  Using the
identity $\det M=\exp(\tr\ln M)$ it is easy to get the usual expression
$\O_1=\tr M^{-1}M'$, where $M'$ is the first derivative of $M$ with
respect to a chemical potential.  Next, using the chain rule
\beq
   \frac{dM^{-1}}{d\mu_f} = -M^{-1}M'M^{-1},
\label{deriv}\eeq
which comes from the identity $MM^{-1}=1$, we recover the relation
$\O_2=\tr (-M^{-1}M'M^{-1}M' + M^{-1}M'')$, where $M''$ is the
second derivative of $M$ with respect to the chemical potential.
Higher operators can be derived by repeated application of the chain
rule with eq.\ (\ref{deriv}), and involve higher derivatives of $M$,
which we write as $M^{(n)}$ (a systematic method for doing this is
given in the appendix). In particular, 
\beqa
\nonumber
   \O_3 &=& \tr\bigg[2(M^{-1}M')^3 - 3M^{-1}M''M^{-1}M' + M^{-1}M^{(3)}\bigg]\\
\nonumber
   \O_4 &=& \tr\bigg[-6(M^{-1}M')^4 - 3(M^{-1}M'')^2 + 12M^{-1}M''(M^{-1}M')^2\bigg.\\
       &&\qquad\qquad\qquad\bigg.
         - 4M^{-1}M^{(3)}M^{-1}M' + M^{-1}M^{(4)}\bigg].
\label{op34}\eeqa
This completes the lattice definitions of the operators.

Before we proceed to evaluate them and extract the non-linear
susceptibilities, we note an ambiguity that arises on the lattice due
to the fact that there is no unique way of putting chemical potential on
the lattice. One can associate a factor $f(a\mu)$ for the propagation of
a quark forward in time by one lattice spacing and a factor $g(a\mu)$
for the propagation of an antiquark. There are exactly four physical
conditions that these two functions must satisfy \cite{chempot}. In the
absence of chemical potential the usual lattice theory must be recovered,
hence $f(0)=g(0)=1$. CP symmetry gives $f(-a\mu)=g(a\mu)$.  Finiteness of
the energy density is guaranteed if $f(a\mu)g(a\mu)=1$. Finally, the
correct continuum limit requires $f'(0)=1$.  These constraints imply the
further relations, $f''(0)=1$ and $f^{(n)}(0) = (-1)^ng^{(n)}(0)$, where
the superscript $n$ on $f$ and $g$ denotes the $n$-th derivative.  All
this guarantees that $n_f$ and $\chi_{fg}$ are prescription independent.

The four conditions above also give relations between the remaining
$f^{(n)}$, such as $f^{(4)}=4f^{(3)}-3$, but do not fix their numerical
values.  Since $\mu$ appears linearly in the continuum Lagrangian,
these higher derivatives are all lattice artifacts.  Any extra
conditions imposed to fix them cannot be physical, and must remain at
the level of prescription.  The usual prescription, $f(a\mu)=\exp(a\mu)$
\cite{haskar}, which we call the HK prescription, gives $f^{(n)}(0)=1$,
but the alternative BG prescription $f(a\mu)=(1+a\mu)/\sqrt(1-a^2\mu^2)$
\cite{bilgav} gives $f^{(3)}(0)=3$ and $f^{(4)}(0)=9$. 

The difference between the two prescriptions can be rather significant. At
any fixed cutoff, one may try to roughly map two prescriptions on to
each other by changing $\mu$ while holding $Z$ fixed by keeping $f(a\mu)$
unchanged.  This gives the relation that for constant physics we must have
\beq
    a\mu_{BG} = \tanh(a\mu_{HK}),
\label{relation}\eeq
where this mapping is for quark chemical potentials.  On $N_t=4$
lattices, the critical end-point for 2+1 flavour QCD has been
determined to be at $T_E=160\pm3.5$ MeV and $\mu_E^{HK}=725\pm3$ MeV
\cite{fodor}. The matching formula of eq.\ (\ref{relation}) then shows
that $\mu_E^{BG}\simeq692$ MeV, and hence the resultant uncertainty
in $\mu_E$ from this source alone is about 11 times larger than the
statistical errors. We next show that this ambiguity vanishes in
the continuum limit in all prescriptions. We also show later (Table
\ref{tb.quench}) that uncertainties of almost 20\% are also expected
from other finite lattice spacing effects even within one prescription,
and lattice spacings of $1/12T_E$ may be required to find $\mu_E$ stable
within statistical error bars.

This freedom of choosing a prescription has specific consequences for the
third and higher derivatives of $M$, and through them for the non-linear
susceptibilities, and hence for $F$, $P$ and all quantities at finite
$\mu$ and $a$.  At $\mu_f=0$, the derivatives of $M$ are related by
\beq
   M^{(n)}=f^{(n)} a^{n-2}M'\ (n{\rm\ odd})\qquad
   M^{(n)}=f^{(n)} a^{n-2}M''\ (n{\rm\ even}).
\label{aleph}\eeq
As a result, $\O_3 = \O_3^{HK} + \Delta
f^{(3)}a^2\O_1$ and $\O_4 = \O_4^{HK} + 4\Delta f^{(3)}a^2\O_2$, where
the superscript HK on an operator denotes its value obtained in the HK
prescription and $\Delta f^{(3)}=f^{(3)}-1$.  Clearly, the prescription
dependence, manifested as a non-vanishing $\Delta f^{(3)}$ at this order,
disappears in the continuum limit, $a\to0$. Since $\langle\O_1\rangle=0$
at $\mu=0$, the prescription dependence of $\langle\O_3\rangle$ is
invisible.  We find that $\chi_{uuud}=\chi_{uuud}^{HK}+\Delta f^{(3)}
(\chi_{ud}/T^2)/N_t^2$. Since $\chi_{ud}$ vanishes within errors, as
we show later, $\chi_{uuud}$ turns out to be effectively prescription
independent.  From the relation for $\O_4$ we find, on varying $N_t$
at fixed $T$,
\beq
   \chi_{uuuu}=\chi_{uuuu}^{HK}+\Delta f^{(3)}
      \left(\frac{\chi_{uu}}{T^2}\right)\left(\frac4{N_t^2}\right).
\label{presc}\eeq
Finally, $\chi_{uudd}$ involves neither $M^{(3)}$ nor $M^{(4)}$,
and hence is prescription independent.  The prescription dependence
of other susceptibilities can be systematically worked out, and it
can be shown exactly as above that they become physical only in the
continuum. Mixed derivatives of $T$ and $\mu$ also have similar behaviour.
If the dependence on $a$ of each susceptibility were known in any scheme,
then one could write down an improved prescription by removing finite $a$
effects systematically.  In other schemes every quantity is potentially
prescription dependent at finite lattice spacing.

\begin{table}\begin{center}\begin{tabular}{l|l||c|c|c|c}
 \hline
 $T/T_c$ & $m_V/T$ & $10^6\chi_{ud}/T^2$ & $10^6\chi_{uuud}$ & $10^4\chi_{uudd}$ & $\mu_*^{HK}/T$ \\
 \hline
 1.0 & 0.2 & 6 (30) & 4 (17) & 7 (1) & 3.20 (3) \\
     & 0.1 & 8 (42) & 7 (33) & 9 (2) & 3.31 (5) \\
     & 0.03 & 11 (84) & 20 (172) & 11 (2) & 3.38 (4) \\
 \hline
 1.5 & 0.2 & -0.3 (423) & -0.7 (116) & 0.107 (3) & 3.73 (1) \\
     & 0.1 & 0.6 (431) & -0.6 (128) & 0.105 (3) & 3.84 (1) \\
     & 0.03 & -0.07 (433) & -0.5 (166) & 0.106 (3) & 3.86 (2) \\
 \hline
 2.0 & 0.2 & 2 (36) & 0.5 (85) & 0.097 (3) & 3.83 (1) \\
     & 0.1 & 2 (36) & 0.5 (89) & 0.098 (3) & 3.87 (1) \\
    & 0.03 & 1 (35) & 0.6 (82) & 0.096 (3) & 3.78 (2) \\
 \hline
 3.0 & 0.2 & 0.6 (19) & 0.1 (5) & 0.032 (2) & 3.87 (1) \\
     & 0.1 & 0.6 (20) & 0.1 (5) & 0.033 (2) & 3.88 (2) \\
     & 0.03 & 0.6 (20) & 0.1 (5) & 0.033 (2) & 3.88 (2) \\
 \hline
\end{tabular}\end{center}
\caption{Results in two flavour QCD with sea quark $m/T_c=0.1$. For
   $T=T_c$ the results are based on 2017 configurations, for $1.5T_c$
   on 370, for $2T_c$ on 126 and for $3T_c$ on 60. At $T_c$ and
   $3T_c$ 100 noise vectors were used. $\chi_{uuuu}$ can be extracted from $\mu_*$
   and $\chi_{uu}$ using eq.\ (\ref{convrg}).}
\label{tb.tdep}\end{table}

After this analysis of lattice artifacts in the Taylor coefficients, we
return to the Taylor expansion itself.  Along the
line $\mu_u=\mu_d=\mu$, the Taylor series expansion of $P$ can be written
in the form
\beq
   \frac{\Delta P}{T^4} = \left(\frac{\chi_{uu}}{T^2}\right)
      \left(\frac\mu T\right)^2 \left[1+\left(\frac{\mu/T}{\mu_*/T}\right)^2
      +\O\left(\frac{\mu^4}{\mu_*^4}\right)\right],
\label{pres}\eeq
where $\Delta P=P(\mu)-P(\mu=0)$, we have neglected $\chi_{uudd}$ in anticipation
of our numerical results (Tables \ref{tb.tdep} and \ref{tb.quench}), and
\beq
   \frac{\mu_*}T= \sqrt{\frac{12\chi_{uu}/T^2}{|\chi_{uuuu}|}}.
\label{convrg}\eeq
For an ideal gas in the continuum, $\chi_{uu}/T^2=1$ and $\chi_{uuuu}=6/\pi^2$,
giving $\mu_*/T=\sqrt2\pi\simeq4.43$. Some remarks are in order---
\begin{enumerate}
\item The series within square brackets in eq.\ (\ref{pres}) is prescription
   dependent at any non-zero lattice spacing, and hence physical values of
   $\Delta P$ can be most reliably extracted by extrapolating each term in
   the series to the continuum.
\item For those values of $\mu$ at which the second or higher order terms in
   the brackets in eq.\ (\ref{pres}) are important, computations of
   $\Delta P/T^4$ on lattices with finite $N_t$ are necessarily prescription
   dependent. Since $F=-PV$, the same is evidently true for all other physical
   quantities, including the energy density.
   From eqs.\ (\ref{presc} and \ref{convrg}), it is clear that
   the prescription dependence of the quadratic term is $(\mu_*/T)^2/3N_t^2$.
   For $N_t=4$ this can be as large as 33\% (see Table \ref{tb.quench}).
\item If the series in eq.\ (\ref{pres}) is well behaved, \ie, sixth and
   higher order susceptibilities are not much larger than $\chi_{uuuu}$,
   then this expansion must be well approximated by the leading term for
   $\mu\ll\mu_*$ in every prescription, and hence be effectively independent
   of prescription \cite{footmub}. Other finite lattice spacing effects may
   still exist.
\item The series expansion must fail to converge in the vicinity of
   a phase transition; therefore estimates of $(T_E,\mu_E)$ on finite lattices
   must be prescription dependent, as we have already estimated. Computation
   of the continuum limit of several terms in the double series for $F(T,\mu)$
   may allow us to use series extrapolation methods, such as Pad\'e approximants
   or estimates of radius of convergence \cite{domb}, to identify $(T_E,\mu_E)$
   in the continuum limit.
\end{enumerate}

\begin{table}[phtb]\begin{center}\begin{tabular}{l|r||c|c|c|c|c}
 \hline
 $T/T_c$ & $N_t$ & $10^6\chi_{ud}/T^2$ & $10^6\chi_{uuud}$ & $\chi_{uuuu}$ & $\mu_*^{HK}/T$ \\
 \hline
 1.5 & 4 & 2 (28) & -0.7 (56) & 1.48 (2) & 3.81 (2) \\
     & 8 & 0.2 (15) & 0.2 (13) & 0.70 (1) & 4.36 (4) \\
     & 10 & -0.4 (77) & 0.04 (64) & 0.61 (2) & 4.47 (4) \\
     & 12 & -0.5 (5) & 0.00 (30) & 0.56 (1) & 4.55 (4) \\
     & 14 & 0.9 (58) & 0.00 (24) & 0.53 (1) & 4.56 (4) \\
 \hline
     & $\infty$ & --- & --- & 0.45 (1) & 4.67 (4) \\
 \hline
 2.0 & 6 & 0.2 (67) & 0.2 (10) & 1.01 (1) & 4.11 (1) \\
     & 8 & -0.3 (115) & 0.1 (13) & 0.74 (1) & 4.32 (5) \\
     & 10 & -0.3 (76) & 0.007 (49) & 6.37 (3) & 4.45 (3) \\
     & 12 & 0.0 (57) & 0.00 (34) & 0.58 (1) & 4.56 (3) \\
     & 14 & -0.2 (43) & 0.00 (17) & 0.56 (1) & 4.59 (4) \\
 \hline
     & $\infty$ & --- & --- & 0.49 (3) & 4.76 (4) \\
 \hline
 3.0 & 4 & 2 (25) & 0.8 (44) & 1.54 (1) & 3.85 (1) \\
     & 8 & 2 (4) & 0.1 (4) & 0.79 (2) & 4.25 (5) \\
     & 10 & -0.6 (14) & -0.04 (11) & 0.66 (1) & 4.40 (3) \\
     & 12 & -0.1 (17) & -0.02 (7) & 0.61 (1) & 4.48 (4) \\
     & 14 & -0.2 (8) & 0.00 (3) & 0.58 (1) & 4.51 (2) \\
 \hline
     & $\infty$ & --- & --- & 0.496 (1) & 4.62 (1) \\
 \hline
\end{tabular}\end{center}
\caption{Results in quenched QCD with $m_v/T_c=0.1$. Quadratic extrapolations
   to the continuum limit, $N_t=\infty$, from the last three points, are shown.
   $\mu_*$ and $\chi_{uuuu}$ are related by eq.\ (\ref{convrg}).}
\label{tb.quench}\end{table}

We turn now to our numerical simulations.  For dynamical sea quark mass
$m/T_c=0.1$ we studied the higher order susceptibilities at $T=T_c$ on a
$4\times10^3$ lattice, $1.5T_c$ and $2T_c$ on $4\times12^3$ lattices, and
$3T_c$ on a $4\times14^3$ lattice.  All the simulations were performed
using the hybrid R-algorithm \cite{ralg} with molecular dynamics
trajectories integrated over one unit of MD time using a leap-frog
algorithm with time step of 0.01 units. At $T_c$ autocorrelations of
the Wilson line and the quark condensate were found to be between 150
and 250 trajectories. With over 2000 saved configurations separated by
10 trajectories each, this gave the equivalent of about 100 independent
configurations. For $T>T_c$ the autocorrelations were all less than 10
trajectories, and hence all the saved configurations can be considered
statistically independent.

Quark number susceptibilities were evaluated in the HK prescription on
stored configurations using valence quark masses $m_V/T_c=0.2$, 0.1 and
0.03. The smallest valence quark mass is chosen such that the ratio of
the ($T=0$) rho and pion masses reaches its physical value $0.2$ at the
lattice spacing $a=1/4T_c$. All quark-line disconnected diagrams of the
kind needed for these measurements are evaluated using a straightforward
extension of the stochastic method given for $\chi_{ud}$ in \cite{conti}
using 10 to 100 noise vectors per configuration \cite{noise}. Our
results for the non-linear susceptibilities which do not vanish by
symmetry are shown in Table \ref{tb.tdep}.  It is clear that of these
only $\chi_{uudd}$ and $\chi_{uuuu}$, are non-zero with statistical
significance.  Comparing them to computations with sea quark mass
$m/T_c=0.2$ and various volumes, we concluded that they are free of
sea quark mass and finite volume effects.  Also note the stability in
physical quantities as $m_v/T_c$ decreases from 0.1 to 0.03.

With present day computer resources the continuum limit is hard to
take in QCD with dynamical quarks. To investigate this limit we have
evaluated the same quantities in quenched QCD for $T\ge1.5T_c$ where the
difference in the order of transitions is immaterial \cite{pushan}. The
run parameters are exactly as in \cite{conti}. Our results are shown
in Table \ref{tb.quench}.  These results show that there is over 20\%
movement in $\mu_*$ when going from $N_t=4$ to the continuum within
a fixed prescription. Since $\mu_*$ is an estimate of the radius of
convergence of the Taylor expansion at the fourth order, it implies that
the estimate of the end-point, $\mu_E$, may shift upward by about 20\%
due to finite size effects even inside the HK scheme.  $\chi_{uudd}$
remains significantly non-zero on all the lattices, and there is some
evidence that it becomes either zero or marginally negative in the
continuum \cite{footnz}. We shall present more detailed studies in
the future.  Finally. the results for $N_t=4$ are very similar in the
quenched and dynamical theories, leading us to believe that the continuum
limits will also be close.

\begin{figure}[tbh]
   \begin{center}\scalebox{0.75}{\includegraphics{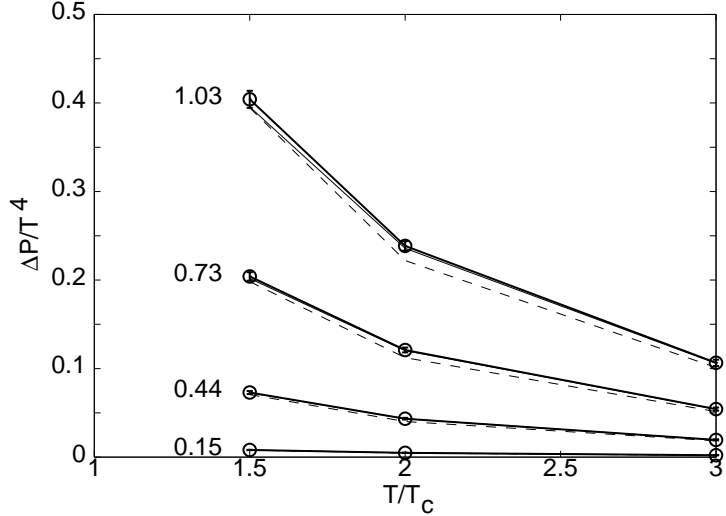}}\end{center}
   \caption{$\Delta P/T^4$ as a function of $T/T_c$ for the values of
      $\mu/T_c$ shown. Continuum results correct to $\O(\mu^4)$ (full lines)
      and $\O(\mu^2)$ (dotted lines) are shown. $N_t=4$ results, in the HK
      prescription, correct to $\O(\mu^4)$ and multiplied by 0.47 to compensate
      for finite $a$ effects in $\chi_{uu}$ are shown with dashed lines.}
\label{fg.pres}\end{figure}

$\Delta P/T^4$ obtained in quenched QCD, using values of $\chi_{uu}$
from \cite{conti} and $\mu_*/T$ obtained here, are shown in Figure
\ref{fg.pres}.  At RHIC it is seen that $\mu/T_c=0.06\ll0.15$, which
implies that $\Delta P/T^4$ is negligible.  In terms of dimensionless
variables, the results in quenched and dynamical QCD are not expected
to differ by more than 5--10\% \cite{fod2}. For $\mu/T_c\simeq0.45$,
relevant to SPS energies, the effects of $\mu>0$ are more significant,
but can still be reliably extracted using only the leading term
of eq.\ (\ref{pres}). In this whole range of $\mu/T_c$ the results
of \cite{fod2}, including a correction for finite lattice spacing
artifacts in the evaluation of $\chi_{uu}$ at $N_t=4$, are the same as
our continuum results, and both are dominated by the leading term of
eq.\ (\ref{pres}). Our computations show that for $\mu\ge2T_c$, higher
order terms become significant for the continuum limit. As a result, at
these chemical potentials, reweighting on $N_t=4$ lattices, even after
correcting for finite $a$ effects in $\chi_{uu}$, are quite different
from the continuum values.

In conclusion, we have studied non-linear susceptibilities and shown
that they are prescription dependent at finite lattice spacing. We
have found the continuum limit of these quantities in quenched QCD, and
thereby removed these artifacts. This allows us to compute the finite
chemical potential corrections to the pressure relevant to RHIC and SPS
experiments.  For $a=1/4T$ the numerical results for QCD with and without
dynamical quarks are similar, and we find the continuum limit of some
of these quantities in the quenched theory. It would be interesting to
compare them with perturbation theory.  We have argued that the critical
end point $(T_E,\mu_E)$ evaluated at $N_t=4$ is uncertain by more than
10 times the statistical errors. As a result, a continuum extrapolation
is required to obtain the physical value of the end point. This may be
possible with the computation of several non-linear susceptibilities.

We would like to thank J.-P.\ Blaizot for discussions.

\appendix\section{Lattice operators}

In this appendix we work in lattice units, \ie, we choose the lattice
spacing to be unity. We introduce the compact notation
\beq
   \tr\left[\bigg(M^{-1}M^{(p_1)}\bigg)^{n_1} \bigg(M^{-1}M^{(p_2)}\bigg)^{n_2}\cdots\right]
    = (n_1\cdot p_1\oplus n_2\cdot p_2\oplus \cdots),
\eeq
and further write $(1\cdot p)$ as $(p)$. 
Since the trace allows only cyclic permutations, therefore
\beq
   (a\oplus b\oplus c) = (c\oplus a\oplus b) \ne (b\oplus a\oplus c),
\eeq
\ie, the `addition' (represented by $\oplus$) is not commutative. `Multiplication' (denoted by the dot)
is distributive over addition, subject to restrictions due to non-commutativity, \ie,
\beqa
\nonumber
   (n\cdot p\oplus m\cdot p) &=& ((n+m)\cdot p),\\
   (n\cdot p\oplus m\cdot p'\oplus l\cdot p)&=&((n+l)\cdot p\oplus m\cdot p'),
\eeqa
but no simplification is possible for $(n\cdot p\oplus m\cdot p'\oplus l\cdot p\oplus\cdots)$.
Traces can be added, \ie,
\beq
   a (n\cdot p) + b (n\cdot p) = (a+b) (n\cdot p).
\eeq
The point of all this is to simplify the taking of derivatives. These are easy to write---
\beq
   (n\cdot p)' = -n (1\oplus n\cdot p)+ n((n-1)\cdot p\oplus (p+1)).
\label{der}\eeq
The operation of taking derivatives is linear over the `addition' $\oplus$,
since this is just the rule for taking derivatives of products.

We have the first examples
\beq
   \O_1 = (1),\qquad\O_2 = -(2\cdot1)+(2).
\label{o12}\eeq
Then, the remaining known ones are obtained simply by applying the rules again. Since
$(2\cdot1)'=-2(3\cdot1)+2(1\oplus 2)$ and $(2)'=-(1\oplus 2)+(3)$, we first obtain
the relation in eq.\ (\ref{op34}),
\beq
   \O_3 = 2(3\cdot1)-3(1\oplus 2)+(3).
\label{o3}\eeq

At the fourth order we need the derivatives
\beqa
\nonumber && (3\cdot1)'=-3(4\cdot1)+3(2\cdot1\oplus 2),\\
\nonumber && (1\oplus 2)'=-2(2\cdot1\oplus 2)+(2\cdot2)+(1\oplus 3),\\
          && (3)'=-(1\oplus 3)+(4).
\label{der4}\eeqa
As a consequence of the general rule in eq.\ (\ref{der}), the coefficients
sum up to zero. This is a consequence of the rule for derivatives
in eq.\ (\ref{der}). Also note that each operator, $(\cdots\oplus
n_i\cdot p_i\oplus \cdots)$, which contributes to $\O_n$ must satisfy
the constraint $\sum n_ip_i=n$. The expressions in eq.\ (\ref{der4})
give the result of eq.\ (\ref{op34}),
\beq
   \O_4 = -6(4\cdot1)+12(2\cdot1\oplus 2)-3(2\cdot2)-4(1\oplus 3)+(4).
\label{o4}\eeq
For each $\O_n$ for $n\ge2$, the sum of the coefficients is zero, as can
be proved by induction from eq.\ (\ref{der}).

Using these rules higher order derivatives, needed for the higher order
susceptibilities, can be easily written down. Since these manipulations
are simple rules for rewriting expressions, not only are they easy to
automate inside standard algebra packages, but can even be readily
implemented as macros in text editors such as sed or emacs.


\begin{thebibliography}{99}
\bibitem{footz}
   $Z$ is, of course, real and positive even for $\mu>0$. In the quenched
   theory valence quarks exist and $\mu$ acts on them.
\bibitem{footdet}
   The determinant for each flavour is obtained as usual by taking the fourth
   root of each staggered quark determinant. As a result, there is a factor
   $1/4$ for each quark loop evaluated with staggered quarks.
\bibitem{footmub}
   Interestingly, at finite isospin density ($\mu_u=-\mu_d=\mu_3/2$) the
   expression for $\Delta P/T^4$ is also given by eq.\ (\ref{pres}). For
   imaginary $\mu_B$, $\Delta P/T^4$ is negative and the higher order
   terms in eq.\ (\ref{pres}) alternate in sign.
\bibitem{noise}
   Our results are stable under change of such algorithmic parameters.
\end{thebibliography}
\end{document}